# Trading Signals In VIX Futures*

M. Avellaneda†, T. N. Lǐ‡, A. Papanicolaou§, G. Wáng¶

November 22, 2021

**Abstract**

We propose a new approach for trading VIX futures. We assume that the term structure of VIX futures follows a Markov model. Our trading strategy selects a position in VIX futures by maximizing the expected utility for a day-ahead horizon given the current shape and level of the term structure. Computationally, we model the functional dependence between the VIX futures curve, the VIX futures positions, and the expected utility as a deep neural network with five hidden layers. Out-of-sample backtests of the VIX futures trading strategy suggest that this approach gives rise to reasonable portfolio performance, and to positions in which the investor will be either long or short VIX futures contracts depending on the market environment.

## 1 Introduction

The shape of the VIX futures curve is informative if it shows a shape that is likely to persist for only a short period of time. In this situation, there may be a simple VIX futures trade that will produce profits when the curve reverts to a more typical shape. For example, if the curve has a hump then there may be a long-short VIX futures position, or a calendar spread, with zero entry cost, which will pay a positive amount when the curve reverts to contango. Ideally, such a reversion will happen quickly so that the trade generates a profit with near certainty. In practice there is some risk because most trades involve non-zero probability of losses. Nevertheless, over long-term horizons with multiple trading opportunities, losses can be diminished if trading strategies are constructed to optimize the expected value of a suitable utility function. VIX futures are a good choice for such trading strategies because their curves have a propensity to quickly revert to contango, which allows for fast turnaround before the next trading opportunity.[1]

We use a stationary VIX futures curve model, as done in Avellaneda and Papanicolaou (2019), to generate day-ahead scenarios of VIX futures. Let $U(\cdot)$ denote a chosen utility function. A trading signal is the optimal trading action that maximizes expected utility under the probability distribution of the model,

$$\boldsymbol{a}(\boldsymbol{x}) = \underset{\boldsymbol{a}\in\mathcal{A}}{\arg\max}\ \mathbb{E}\left[U\left(\boldsymbol{R}_{t+1}(\boldsymbol{a})\right)|\boldsymbol{X}_t = \boldsymbol{x}\right] , \tag{1.1}$$

where $t$ denotes time, $\mathbb{E}$ denotes expected value, and where

$$\begin{aligned} \boldsymbol{X}_t &= \text{VIX futures curve at time } t\text{, vector valued,} \\ \mathcal{A} &= \text{a set of possible trades/actions } \boldsymbol{a}, \text{vector valued,} \\ \boldsymbol{R}_{t+1}(\boldsymbol{a}) &= \text{change in position from time } t \text{ to } t+1 \text{ if action } \boldsymbol{a} \in \mathcal{A} \text{ is taken.} \end{aligned}$$

*The authors would like to thank Brian Healy and Xunyang Wu for their feedback and support.
†Department of Mathematics, New York University. 251 Mercer Street, New York, NY, 10012. *avellane@cims.nyu.edu*
‡Department of Mathematics, New York University. 251 Mercer Street, New York, NY, 10012. *thomli@cims.nyu.edu*
§Department of Mathematics, North Carolina State University. Campus Box 8205, Raleigh, NC 27695. *apapani@ncsu.edu*. The author is partially supported by NSF grant DMS-1907518.
¶Department of Mathematics, Columbia University. 2990 Broadway New York, NY 10027. *gw2376@columbia.edu*
[1] Here, "quickly" means relative to other curves such as crude oil or treasuries.

The action space $\mathcal{A}$ consists of various positions in VIX futures, and $\boldsymbol{R}_{t+1}(\boldsymbol{a})$ is a function of the action $\boldsymbol{a}$ and the transition occurring in the VIX futures curve,

$$(\boldsymbol{X}_t, \boldsymbol{X}_{t+1},\ \boldsymbol{a}) \mapsto \boldsymbol{R}_{t+1}(\boldsymbol{a}) \ .$$

We take $\mathcal{A}$ to be a finite set of trades that are predetermined, and we assume that the transition distribution for $\boldsymbol{X}_t$ is also given. We estimate the expected value in equation (1.1) using a deep neural network, see Goodfellow et al. (2016). Historical VIX futures data are applied to estimate the parameters for the model of $\boldsymbol{X}_t$, and then the neural network is trained using simulated data generated by this estimated model. In our model, the most likely curve is a contango, and all other curve shapes will revert toward this most likely state. To illustrate, Figure 1.1 shows a contango and a backwardation curve of VIX futures. We construct a trading signal by solving the optimization problem (1.1) with $\mathcal{A}$ consisting of four different allocations in one-month and five-month rolling VIX futures strategies (see Section 2.1 where we define these rolling strategies). For most contango curves, the action suggested by the trading signal is to long the one-month strategy and to short the five-month strategy. In backwardation, the suggested trade is to short the one-month and go $2\times$ long the five-month. In backtesting of this trading signal, we find that if transaction costs are not too high, then for a trading period of around 200 days, there can be profits of double-digit percentage and Sharpe ratios significantly higher than one.

## 1.1 Literature of Related Research

The VIX has been the "fear gauge" for the financial markets of the United States since 1993, see Whaley (2000) and Whaley (2009). Since 2004, the market for VIX futures has made it possible to gain exposure to VIX, and the creation of exchange-traded notes (ETNs) has made it possible to gain exposure with greater ease, see Alexander et al. (2015). The significance of mean reversion and contango in VIX futures and ETNs is analyzed in Avellaneda and Papanicolaou (2019). Mean reversion is also the key assumption in the class of stochastic volatility models driven by stationary factor processes, see for instance Fouque et al. (2000). Historically, volatility models in finance have relied on the Markov property, but recently there has been a trend toward VIX pricing driven by fractional Brownian motion, see Bayer et al. (2016). A Markovianization of the fractional-curve model is achieved by considering the infinite-dimensional futures curve in its entirety, see Euch and Rosenbaum (2018). Foundational concepts in machine learning such as convergence and deep learning extensions can be found in Mohri et al. (2018) and Sutton and Barto (2018). The implementation of high-dimensional learning has been made possible by recent developments in neural network software such as TensorFlow and PyTorch. An example of note is the deep-Q neural network (DQN) algorithm, see Mnih et al. (2015) and Fan et al. (2020). For applications to finance see Aldridge and Avellaneda (2020), Sirignano and Spiliopoulos (2017), Casgrain et al. (2019), and Ruf and Wang (2021). Studies on high-dimensional deep learning have highlighted the improvement in out-of-sample prediction when large neural networks are utilized, see Zhang et al. (2017), Belkin et al. (2018), and Hastie et al. (2022). Evaluation of out-of-sample performance is often done using cross-validation methods, but special care needs to be taken when applying these methods to financial data, see Arlot and Celisse (2010), Arnott et al. (2019), and Ruf and Wang (2020). In particular, with times series data there can be significant auto-correlations, yet cross-validation methods are still applicable so long as the time series are assumed to satisfy some basic assumptions such as zero auto-correlations in the noise process, see Burman and Nolan (1992), Bergmeir and Benítez (2012), and Bergmeir et al. (2018).

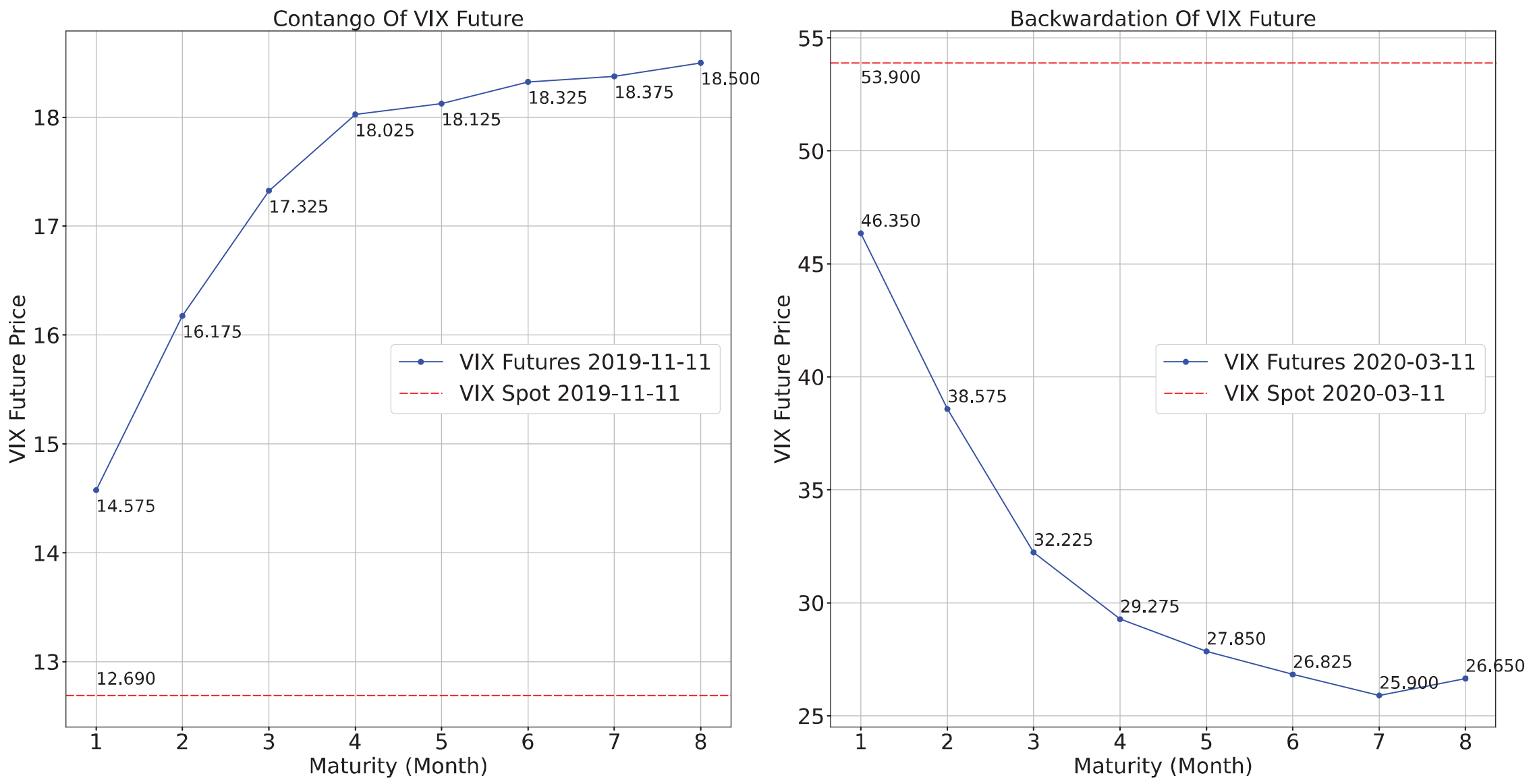

Figure 1.1: The VIX futures' contango curve seen on 2019-11-11 (left) and the backwardation seen on 2020-03-11 (right). A trading signal is constructed based on the value and shape of this curve.

## 1.2 Main Results and Structure of the Article

The focus of this article is on a new method for trading VIX futures, wherein trading signals are the optimal action function given by equation (1.1). We implement this new approach on a variety of utility functions and utilize deep neural networks to estimate the objective in equation (1.1). We conduct cross-validation studies using a $k$-fold procedure. We use historical VIX futures data consisting of end-of-day VIX futures curves from January 2008 to February 2021. In out-of-sample tests we find that trading signals constructed with deep neural networks have the potential to produce reasonable profits and Sharpe ratios. These findings are an indication that VIX futures curves contain useful predictive information for trading, and that deep neural networks are able to filter and apply the relevant information from the curves.

The article is organized as follows: Section 2 introduces the VIX curve model, explains how parameters are estimated, and describes the futures positions that we optimize over; Section 3 presents cross-validation studies of the neural network method on historical VIX futures data – both with and without transaction costs; Section 4 concludes; Appendix A shows a real-time backtest that we conducted with weekly re-training of the neural network from December 28th, 2020 through February 19th, 2021; Appendix A also provides a detailed account of how the outputs of the neural network map to exact trading positions in VIX futures; Appendix B provides metrics for various non-neural network benchmarks.

# 2 A Model for Trading VIX Futures

Let $t$ denote time and let $\text{VIX}_t$ denote the value of VIX on that date. Let $d$ be an integer such that $d+1$ is the number of VIX futures contracts[2], and let $T_1 < T_2 < \cdots < T_{d+1}$ denote the expiration dates of these VIX futures contracts. Let us denote

$$F_t^i := \text{VIX future expiring at time}\, T_i\,, \tag{2.1}$$

where $t = 0,\ 1,\ 2,\ \cdots,\ T_i$ is the current date. A term-structure of constant-maturity VIX futures (CMFs), each with horizon $\theta_i$-many months, for $i = 1,\ 2,\ 3,\ \cdots,\ d$, are constructed as a linear interpolation of the VIX futures,

$$V_t^i := \omega_t^i F_t^i + \left(1 - \omega_t^i\right) F_t^{i+1}\,, \tag{2.2}$$

where $t \leq T_i \leq t + \theta_i \leq T_{i+1}$ and $\omega_t^i = \frac{T_{i+1} - t - \theta_i}{T_{i+1} - T_i}$; note now that $V_t^i$ is defined for all $t$. Note also that $\text{VIX}_t$ is like a zero-horizon CMF. CMFs are preferable for statistical estimation because they do not have non-stationary effects that are caused by contract expiry.

## 2.1 Rolling VIX Futures Strategies

A rolling VIX futures strategy maintains the CMF weights of equation (2.2) for fixed maturity $\theta_i$. For each $i$, we let $I^i$ denote the value of the rolling VIX futures strategy with horizon $\theta_i$, for which returns are given by

$$\frac{\Delta I_t^i}{I_t^i} := \frac{\omega_t^i \Delta F_t^i + \left(1 - \omega_t^i\right) \Delta F_t^{i+1}}{\omega_t^i F_t^i + \left(1 - \omega_t^i\right) F_t^{i+1}} + r\Delta t\,, \tag{2.3}$$

where $\Delta I_t^i = I_{t+1}^i - I_t^i$, $\Delta F_t^i = F_{t+1}^i - F_t^i$, $r \geq 0$ is the interest rate, and $\Delta t = \frac{1}{252}$. Simple calculation leads to an equivalent expression to equation (2.3) in terms of the CMFs,

$$\frac{\Delta I_t^i}{I_t^i} = \left(r + \dot{\omega}_t^i \frac{F_{t+1}^{i+1} - F_{t+1}^i}{V_t^i}\right) \Delta t + \frac{\Delta V_t^i}{V_t^i}\,, \tag{2.4}$$

where $\dot{\omega}_t^i = \frac{\omega_{t+1}^i - \omega_t^i}{\Delta t} < 0$ for all $t < T_i$. The drift term in equation (2.4) contains the quantity referred to as the roll yield,

$$\text{Roll}_{t+1}^i := \dot{\omega}_t^i \frac{F_{t+1}^{i+1} - F_{t+1}^i}{V_t^i}\,,$$

[2]The VIX futures term structure is a collection of VIX futures contracts with nine monthly maturities, and six weekly contracts that are not very liquid.

which we utilize to re-write equation (2.4) as follows,

$$\frac{\Delta I_t^i}{I_t^i} = \left(r + \mathrm{Roll}_{t+1}^i\right)\Delta t + \frac{\Delta V_t^i}{V_t^i}\,. \tag{2.5}$$

From equation (2.5) we see that if $V_t^i$ is a stationary process then the return rate of the $i^{th}$ rolling VIX futures strategy has a most likely value equal to the risk-free rate plus the mode of $\mathrm{Roll}_{t+1}^i$. As shown in Avellaneda and Papanicolaou (2019), the most likely VIX futures curves are contango and the most likely roll yields are negative, which explains why the value of the rolling VIX futures strategies decay.

In the past there have been some attempts to apply statistical arbitrage techniques to VIX. One idea is to use the Engle-Granger test to find co-integrated pairs among rolling VIX futures strategies, see Engle and Granger (1987). For the one-month rolling VIX futures strategy ($\theta =$ one month) and the five-month strategy ($\theta =$ five months), a simple linear regression of one set of returns on the other suggests that we should short the one-month and long $0.9\times$ five-month. However, this is not a good pair to trade because the residual is not stationary; for daily data between 2008 and 2020 the values of these positions do not reject a unit root hypothesis. In addition, historical backtesting shows that these trades have large drawdowns and negative returns at the most inopportune times. Another possibility is to match volatility levels between the one-month and five-month rolling VIX futures portfolios, which suggests a position $1\times$ short the one-month and $2\times$ long the five-month, respectively. This was a popular trade during the decade of 2010, but also had large drawdowns. The conclusion is that allocations in these rolling VIX futures portfolios are useful but there needs to be a rule for deciding when to open and close the trade.

**Remark 2.1** (Exchange Traded Notes)**.** *Rolling VIX futures portfolios represent the underlying redemption value for several VIX ETNs. Such notes are among the more liquid instruments for gaining exposure to VIX, see Alexander et al. (2015). Some of the more liquid ETNs include the iPath VXX (long one-month), the iPath VXZ (long five-month), the VelocityShares TVIX (2× long one-month), and the iPath XIV (short one-month). Trading in these notes can be replicated with trades in the rolling VIX futures strategies. However in practice, replication is not entirely accurate. Firstly, the issuer of a note may have call-back features embedded, which can terminate the note at any time. Secondly, the rolling VIX futures strategy is technically just the redemption value and the notes are free to trade at market value, which means that there may be a slight discrepancy between the ETN's returns and its respective rolling futures formula.*

## 2.2 Vector Auto-Regressive Model

The two main quantities that we consider are the CMFs $\left(V_t^i\right)_{i=0}^d$ and the roll yields $\left(\mathrm{Roll}_t^i\right)_{i=1}^d$. As seen from equation (2.5), these quantities can be used to make short-term predictions on the rolling VIX futures strategies. For example, the roll yield and the anticipated direction of mean reversion could be the basis for a trading strategy that performs well in the long term.

For the $t^{th}$ day of a given time period, the VIX futures curve is described by the following state vector,

$$\boldsymbol{X}_t = \left[\log \mathrm{VIX}_t,\ \log V_t^1,\ \log V_t^2,\ \cdots,\ \log V_t^d,\ \mathrm{Roll}_t^1,\ \mathrm{Roll}_t^2,\ \cdots,\ \mathrm{Roll}_t^d\right]^\top,$$

where all entries of this vector are directly computable from $\left(\text{VIX}_s, F_s^1, F_s^2, \cdots, F_s^{d+1}\right)_{s\leq t}$. Given data at times $t = 1, 2, \cdots, T$, let $\boldsymbol{X}^*$ denote the mode,

$$\boldsymbol{X}^* = \underset{t\leq T}{\text{mode}}\left(\boldsymbol{X}_t\right) ,$$

that is, $\boldsymbol{X}^*$ is the most likely curve, which is illustrated in Figure 2.1. The figure displays the mean of the state given by $\frac{1}{T}\sum_{t=1}^{T}\boldsymbol{X}_t$, and the mode of the state. Statistical analysis in Avellaneda and Papanicolaou (2019) shows that $\boldsymbol{X}_t$ is a stationary stochastic process whose historical time series exhibits a tendency to mean revert towards a contango curve. In its most likely state, the VIX future is around 12%-14%, the long-term VIX future is around 17%-20%, and all in-between CMFs lie on an upward sloping curve.

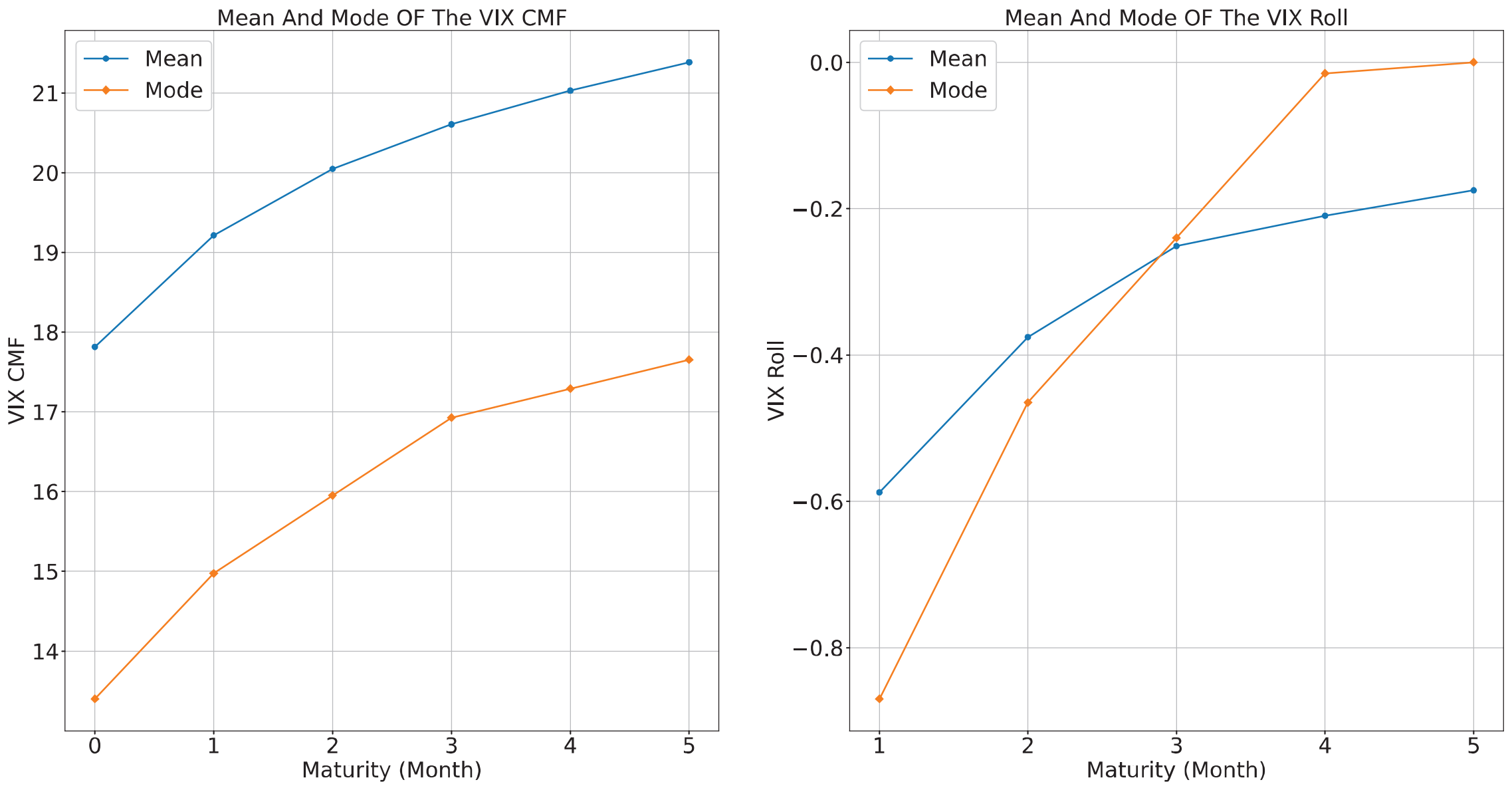

Figure 2.1: The mean and modal curves of VIX CMFs (left) and the mean and modal curves of the roll yields (right). The VIX futures curves are usually in contango, with the possibility of a volatility spike causing an upward skew in the distributions of VIX futures. Therefore, the mean CMF curve is above the modal curve, and a similar relationship appears in the mean and modal curves of the negative roll yields.

We take the state vector $\boldsymbol{X}_t$ for $t = 1, 2, \cdots, T$, center it around the mode, and then place it in a larger matrix

$$\boldsymbol{\psi} = \left[\boldsymbol{X}_1 - \boldsymbol{X}^*, \boldsymbol{X}_2 - \boldsymbol{X}^*, \cdots, \boldsymbol{X}_T - \boldsymbol{X}^*\right] .$$

Note that we are centering around the mode rather than the mean, which we do for robustness because CMFs have a heavy right skew.

The vector auto-regressive (AR) model for the state vector is the following,

$$\boldsymbol{\psi}_{t+1} = \boldsymbol{\mu} + \boldsymbol{A}\boldsymbol{\psi}_t + \boldsymbol{Z}_{t+1} , \tag{2.6}$$

where $\boldsymbol{Z}_t$ is an independent and identically distributed Gaussian random vector with mean zero and covariance $\boldsymbol{\Sigma}$. The least-squares estimator of $\boldsymbol{A}$ is given by

$$\widehat{\boldsymbol{A}} = \left[\sum_{t=1}^{T-1} \left(\boldsymbol{\psi}_{t+1} - \overline{\boldsymbol{\psi}}\right)\left(\boldsymbol{\psi}_{t} - \overline{\boldsymbol{\psi}}\right)^\top\right]\left[\sum_{t=1}^{T-1} \left(\boldsymbol{\psi}_{t} - \overline{\boldsymbol{\psi}}\right)\left(\boldsymbol{\psi}_{t} - \overline{\boldsymbol{\psi}}\right)^\top\right]^{-1},$$
$$\hat{\boldsymbol{\mu}} = \left(\boldsymbol{I} - \widehat{\boldsymbol{A}}\right)\overline{\boldsymbol{\psi}},$$

where $\overline{\boldsymbol{\psi}} = \frac{1}{T}\sum_{t=1}^{T}\boldsymbol{\psi}_t$. The covariance matrix $\boldsymbol{\Sigma}$ can be estimated by

$$\widehat{\boldsymbol{\Sigma}} = \frac{1}{T-1}\sum_{t=1}^{T-1}\widehat{\boldsymbol{Z}}_t\widehat{\boldsymbol{Z}}_t^\top,$$

where $\widehat{\boldsymbol{Z}}_{t+1} = \boldsymbol{\psi}_{t+1} - \hat{\boldsymbol{\mu}} - \widehat{\boldsymbol{A}}\boldsymbol{\psi}_t$.

We can write the returns on the rolling VIX futures strategies from equation (2.5) as

$$\frac{\Delta I_t^i}{I_t^i} = \left(r + \boldsymbol{X}_{t+1}^{d+i}\right)\Delta t + \frac{\exp\left(\boldsymbol{X}_{t+1}^i\right) - \exp\left(\boldsymbol{X}_t^i\right)}{\exp\left(\boldsymbol{X}_t^i\right)}, \qquad \text{for } 1 \le i \le d, \tag{2.7}$$

which will be useful in the sequel where we draw samples from a distribution for $\boldsymbol{X}_t$ and use to simulate trading returns. That is, we will use the vector AR model that is described by equation (2.6) to simulate $\boldsymbol{X}_t$, which we insert into equation (2.7) for computing the returns of rolling VIX futures strategies. Figure 2.2 shows the simulations of the one-month and five-month rolling VIX futures strategies, with each simulation including its respective historical portfolio value.

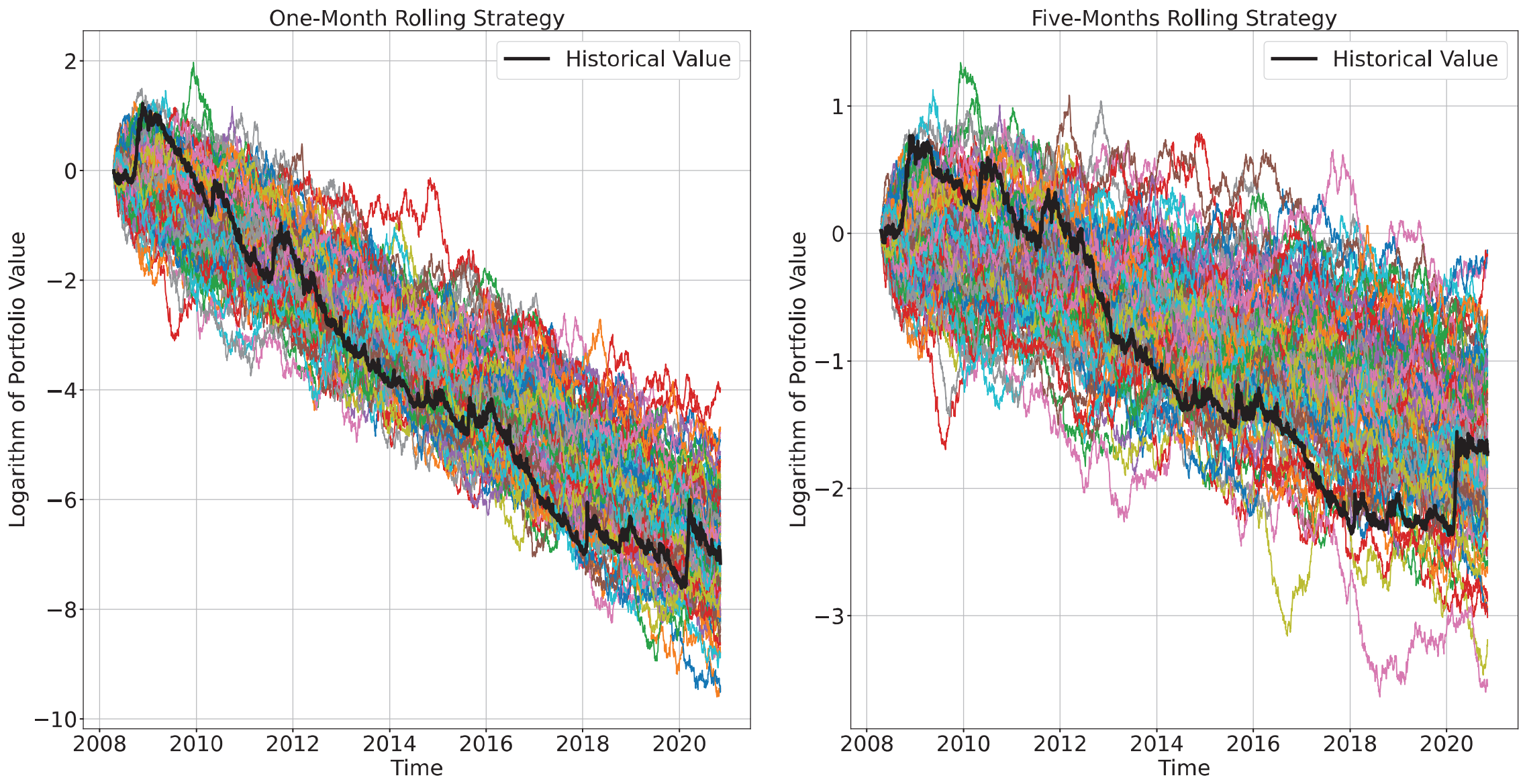

Figure 2.2: Simulations of the one-month rolling VIX futures strategy and the five-month rolling VIX futures strategy, which are generated from the vector AR model in equation (2.6). The dark line in each plot is the historical value of the respective strategy. The declining value in these rolling strategies is studied in Avellaneda and Papanicolaou (2019).

## 2.3 Trading-Signal Construction

We consider the following quantity,

$$\boldsymbol{R}_{t+1}(\boldsymbol{a}) := \sum_i a^i \left( \frac{\Delta I_t^i}{I_t^i} - r\Delta t \right) . \tag{2.8}$$

This represents the profit or loss for a position in rolling VIX futures strategies[3]. Let $\mathcal{A}$ denote the space of admissible actions. An optimal action is determined by maximizing the expected utility,

$$\max_{a \in \mathcal{A}} \mathbb{E} \left[ U(\boldsymbol{R}_{t+1}(\boldsymbol{a})) \middle| \boldsymbol{X}_t = \boldsymbol{x} \right] , \tag{2.9}$$

where the action is decided by the trader at time $t$ immediately before $\boldsymbol{R}_{t+1}(\boldsymbol{a})$ is realized, and where $U(\boldsymbol{R})$ is the utility function.

We denote by $P_t$ the value of the trading-signal portfolio at time $t$, for which returns are computed as

$$\frac{\Delta P_t}{P_t} = \boldsymbol{R}_{t+1}(\boldsymbol{a}(\boldsymbol{X}_t)) + r\Delta t , \tag{2.10}$$

where $\boldsymbol{a}(\boldsymbol{X}_t) = \arg\max_{a \in \mathcal{A}} \mathbb{E}[U(\boldsymbol{R}_{t+1}(\boldsymbol{a})) | \boldsymbol{X}_t]$. In testing, we use the time series of $P_t$ to compute performance metrics, such as profit percentages and Sharpe ratios.

# 3 Computing the Trading Signals with Historical Data

We carry out the method described in Section 2 on historical VIX futures data. Our data is daily, beginning April $14^{th}$ of 2008 and going until November $6^{th}$ of 2020, and consists of one-month, two-month, three-month, fourth-month, five-month, and six-month VIX futures, in other words, $i = 1, 2, \cdots, 6$ and $d = 5$ in equation (2.1). The data is downloadable from the VIX Central website.[4] Using these data, we construct the time series of VIX CMFs and VIX rolls as given by equation (2.2) and equation (2.4), respectively. We take the weights $\omega^i$ that appear in equation (2.2) to be $\omega^i \equiv \omega$ for all $i$ such that there is 100% in the front-month contract as soon as the prior future matures, and then 0% in this front-month at the next maturity date. We analyze the time series of portfolio value utilizing the following performance metrics: annualized expected rate of return denoted by $\mathbb{E}[\boldsymbol{R}_t(\boldsymbol{a}(\boldsymbol{X}_t))]$, volatility denoted by $\mathrm{std}[\boldsymbol{R}_t(\boldsymbol{a}(\boldsymbol{X}_t))]$, trading profit, Sharpe ratio[5], and maximum drawdown.

A standard procedure for in-sample training and out-of-sample testing is straightforward: divide the data into two blocks, with the first block designated for in-sample training, and the second block designated for out-of-sample testing. More specifically, we take the VIX futures curves from April $14^{th}$ of 2008 to August 7th of 2019 for in-sample training, and then utilize the remaining curves from August $8^{th}$ of 2019 to November $5^{th}$ of 2020 for out-of-sample testing. But this out-of-sample test is based on a single portfolio run, which means that good performance could be attributable to luck. Therefore, to make full usage of the data, we apply the method of the $k$-fold cross-validation.

[3] Of course, every strategy of rolling VIX futures is equivalent to an allocation of futures contract (see equation (2.3)).

[4] The VIX Central website: `http://vixcentral.com`

[5] We compute the Sharpe ratios by annualized excess return being divided by annualized standard deviation, where the annualized excess return is $\left[\prod_{t=1}^{T} (1 + \boldsymbol{R}_t(\boldsymbol{a}(\boldsymbol{X}_t)))\right]^{252/T} - (1 + r)$, and annualized standard deviation is $\mathrm{std}[\boldsymbol{R}_t(\boldsymbol{a}(\boldsymbol{X}_t))]\sqrt{252}$.

We divide the data into $k = 10$ folds, each with 316 or 317 days, and then utilize these folds to conduct ten separate in-sample trainings and out-of-sample testings. More specifically, we train on a configuration of nine folds, and then upon the remaining fold we conduct an out-of-sample test, see chapter four of Mohri et al. (2018) for details on $k$-fold cross validation. Table 3.1 gives the precise demarcation dates for the folds. When we paste non-contiguous folds, we exclude the *pasting outlier* when estimating the vector AR model (2.6). For example, in order to out-of-sample test on fold #5, we need to paste fold #4 to fold #6 for training, and in doing so we make sure to exclude the data point at the jump from fold #4 to #6.

| Fold # | Time Interval | Fold # | Time Interval |
|---|---|---|---|
| 0 | 2008-04-16 to 2009-07-17 | 5 | 2014-07-31 to 2015-10-29 |
| 1 | 2009-07-20 to 2010-10-19 | 6 | 2015-10-30 to 2017-02-01 |
| 2 | 2010-10-20 to 2012-01-23 | 7 | 2017-02-02 to 2018-05-04 |
| 3 | 2012-01-24 to 2013-04-29 | 8 | 2018-05-07 to 2019-08-07 |
| 4 | 2013-04-30 to 2014-07-30 | 9 | 2019-08-08 to 2020-11-05 |

Table 3.1: The start date and end date for each of the ten backtesting folds in the $k$-fold cross validation.

Our approach is to utilize the training data to estimate the parameters of the vector AR model (2.6) proposed in Section 2.2, and then to draw samples from the vector AR model to train the neural network. The neural network is an approximation of the functional form of $\mathbb{E}\left[U\left(\boldsymbol{R}_{t+1}\left(\boldsymbol{a}\right)\right)|\boldsymbol{X}_t\right]$, see Cybenko (1989) and Pinkus (1999), for each action $\boldsymbol{a}$ in the action space,

$$\mathcal{A} = \Big\{ (0,\ 0)\,,\ (-1,\ 1)\,,\ (-1,\ 2)\,,\ (1,\ -1)\,,\ (1,\ -2) \Big\}\,, \tag{3.1}$$

where the individual actions are

$$\begin{aligned} (0,\ 0) &= \text{no trade}\,,\\ (-1,\ 1) &= \text{short } I^1 \text{ and long } I^5\,,\\ (-1,\ 2) &= \text{short } I^1 \text{ and } 2{\times}\text{long } I^5\,,\\ (1,\ -1) &= \text{long } I^1 \text{ and short } I^5\,,\\ (1,\ -2) &= \text{long } I^1 \text{ and } 2{\times}\text{short } I^5\,, \end{aligned}$$

and where $I^1$ and $I^5$ denote the one-month and the five-month rolling VIX futures strategies, respectively, as defined by equation (2.5) in Section 2.1 .

The $k$-fold cross validation described above is susceptible to data leakage because utilizing a vector AR model implies some dependence between folds, see Arnott et al. (2019), Ruf and Wang (2020). In the literature, it is argued that cross validation methods can effective for auto-regressive models when noise is uncorrelated, see Arlot and Celisse (2010), Bergmeir and Benítez (2012), Bergmeir et al. (2018), Burman and Nolan (1992), and Cerqueira et al. (2020). To test for data leakage in our cross-validation studies, we re-run our cross-validation studies utilizing the non-adjacent block configurations that are proposed in Bergmeir et al. (2018). More specifically, we in-sample train the model utilizing data from folds #2 through #8, and then out-of-sample test the model utilizing data of fold #0; we in-sample train the model utilizing data from folds #3 through #9, and then out-of-sample test the model utilizing data of fold #1; we in-sample train the model utilizing data from fold #0 and folds #4 through #9, and then out-of-sample test the model utilizing data of fold #2; and etc. The purpose for doing the $k$-fold cross-validations with this configurations is to eliminate the contiguous training folds that may have information about the testing fold. However, when we re-run with these non-contiguous configurations, we observe almost no difference compared with the numbers resulting from standard $k$-fold cross validation.

## 3.1 Neural Network Approach

For general concave utility functions, there is not an explicit calculation for the expected utility $\mathbb{E}\left[U\left(\boldsymbol{R}_{t+1}\left(\boldsymbol{a}\right)\right)|\boldsymbol{X}_t\right]$. Therefore, we use a neural network to find an approximating function. The architecture of the neural network that we implement is a deep feed-forward neural network (DFN), as described in Goodfellow et al. (2016). For a discrete set of actions $\mathcal{A} = \{\boldsymbol{a}_{1,} \boldsymbol{a}_{2,} \cdots, \boldsymbol{a}_p\}$, the universal approximation theorem, see Cybenko (1989) and Pinkus (1999), is a mathematical theorem to ensure that DFN is an effective way to estimate the nonlinear mapping

$$\boldsymbol{X}_t \mapsto \left[Q\left(\boldsymbol{X}_t,\, \boldsymbol{a}_1\right), Q\left(\boldsymbol{X}_t,\, \boldsymbol{a}_2\right), \cdots, Q\left(\boldsymbol{X}_t,\, \boldsymbol{a}_p\right)\right]^\top,$$

where $Q\left(\boldsymbol{x}, \boldsymbol{a}_j\right)$ is optimized to approximate $\mathbb{E}\left[U\left(\boldsymbol{R}_{t+1}\left(\boldsymbol{a}_j\right)\right)|\boldsymbol{X}_t = \boldsymbol{x}\right]$. Our approach is to sample $\boldsymbol{X}_t$ from the vector AR model (2.6) that is proposed in Section 2.2, and then use these samples to train the neural network, and finally perform $k$-fold cross validation to test out-of-sample performance of the optimal neural-network trading actions.

The DFN we use has the specifications depicted in Figure 3.1. We utilized a dense connective structure between layers, in other words, all layers have neurons that are fully connected with the neurons in the previous layer. It has eleven neurons on the input layer, each of which represents an element $\boldsymbol{X}_t^i$ for a given $i$. The number of neurons on the output layer is five, which represents the five actions in the action space $\mathcal{A}$. We set the number of the hidden layers to five, each of them containing $J = 50 \times 11$ neurons. In in-sample training, we generate $10^5$ days of data and run the back-propagation for 15 epochs with a batch size of 160. We used a tableau method to determine the number of hidden layers and the number of neurons per layer. The results of this tableau show that portfolio performance has noticeable decline when we use a dense DFN with too few neurons and layers (e.g., 2 layers with only 20 neurons per layer), and also performs poorly when we use a dense DFN with too many layers and (e.g., 8 layers and 1000 neurons per layer).

In the dense DFN, we choose the activation function $f\left(x\right)$ to be the Parametric Rectified Linear Unit (PReLU) function,

$$f\left(x\right) = \begin{cases} x & \text{for} \quad x \geq 0 \\ \alpha x & \text{for} \quad x < 0\,, \end{cases} \tag{3.2}$$

where $\alpha > 0$ with $\alpha = 0.1$. For all results that we present, we take the PReLU activation function for both the hidden layers and the output layer of the DFN. We repeat all tests using hyperbolic tangent activation function $f\left(x\right) = \tanh\left(x\right)$ and linear activation function $f\left(x\right) = wx + b$ for the output layer, but the results from PReLU are slightly better.

Given the neurons, the layers, and the activation function, the underlying structure of $Q$ :

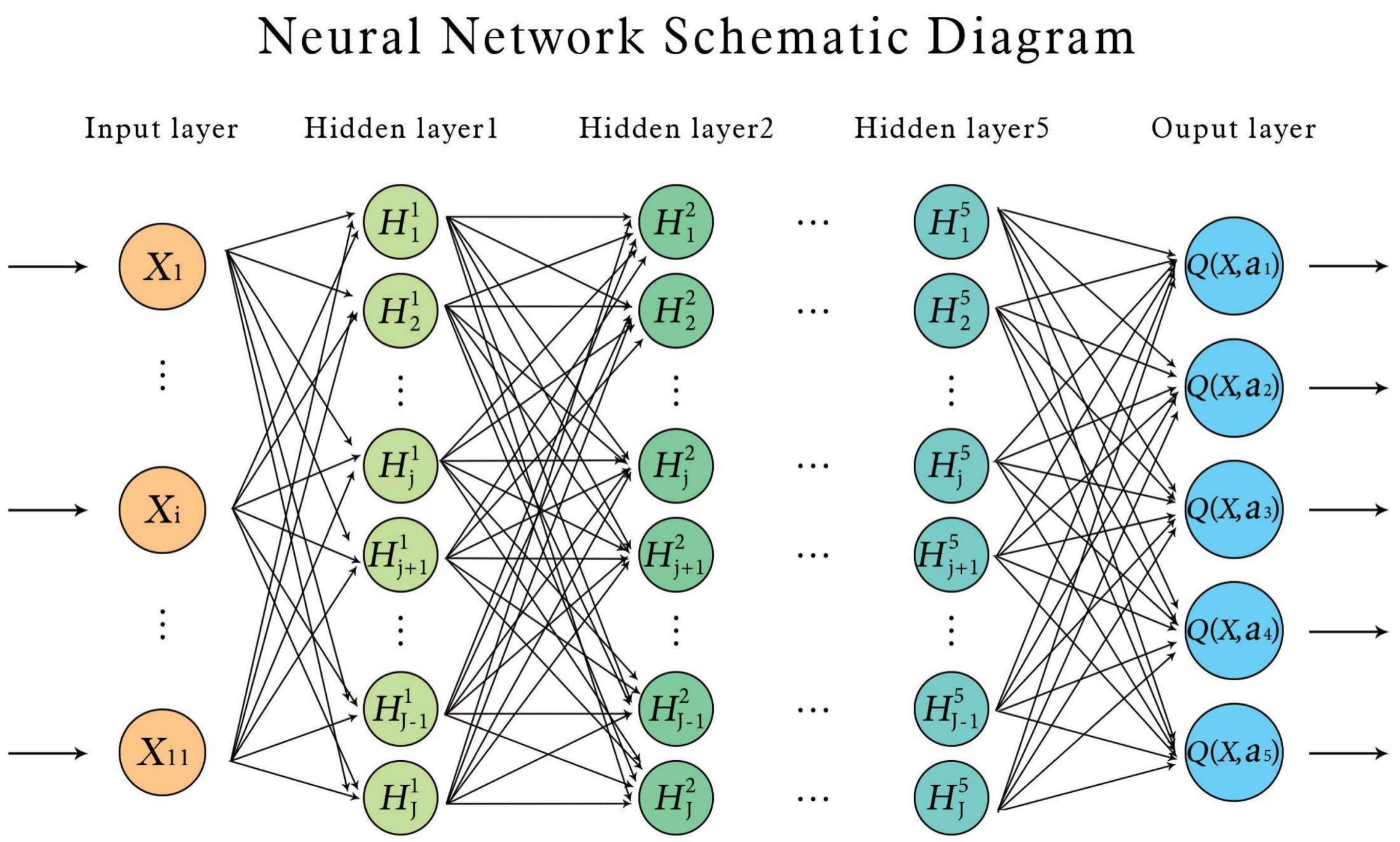

Figure 3.1: Schematic diagram of the deep neural network. In our studies we took $J = 550$, i.e., 550 neurons in each hidden layer.

$\mathbb{R}^{11} \to \mathbb{R}^5$ is

$$\begin{pmatrix} Q(\boldsymbol{X}_t, \boldsymbol{a}_0) \\ Q(\boldsymbol{X}_t, \boldsymbol{a}_1) \\ Q(\boldsymbol{X}_t, \boldsymbol{a}_2) \\ Q(\boldsymbol{X}_t, \boldsymbol{a}_3) \\ Q(\boldsymbol{X}_t, \boldsymbol{a}_4) \end{pmatrix} = f\left(\boldsymbol{W}_6^\top f\left(\cdots f\left(\boldsymbol{W}_2^\top f\left(\boldsymbol{W}_1^\top \boldsymbol{X}_t + \boldsymbol{b}_1\right) + \boldsymbol{b}_2\right) \cdots\right) + \boldsymbol{b}_6\right), \tag{3.3}$$

where $\boldsymbol{W}_\ell$, $\ell = 1, 2, \cdots, 6$ is a matrix of weights connecting the neurons on the $(\ell - 1)^{th}$ layer to the $\ell^{th}$ layer, and $\boldsymbol{b}_\ell$, $\ell = 1, 2, \cdots, 6$ is a vector of biasing values for each layer; here subscript 6 represents the terminal output layer.

We train the dense DFN using samples that are drawn from the vector AR model (2.6) of Section 2.2 with Gaussian noise. We generate independent and identically distributed samples $\boldsymbol{X}_0^{(i)}$ for $i = 1, 2, \cdots, N$ from the stationary distribution of the vector AR model (2.6). We take $N = 10^5$. For each $\boldsymbol{X}_0^{(i)}$, we simulate a batch of one-step forward samples to approximate the conditional expected utility, which we label as $\boldsymbol{R}_1^{(i,i')}(\boldsymbol{a})$ for $i' = 1, 2, \cdots, M$ for each $\boldsymbol{a} \in \mathcal{A}$. We took $M = 300$. We then fit the dense DFN to the sample averages by minimizing the quadratic loss function with respect to the parameters $\boldsymbol{W}$ and $\boldsymbol{b}$,

$$\min_{\boldsymbol{W},\boldsymbol{b}} \sum_{a \in \mathcal{A}} \frac{1}{N} \sum_{i=1}^{N} \left( Q\left(\boldsymbol{X}_0^{(i)}, \boldsymbol{a}\right) - \frac{1}{M} \sum_{i'=1}^{M} U\left(\boldsymbol{R}_1^{(i,i')}(\boldsymbol{a})\right) \right)^2. \tag{3.4}$$

After training, the optimally fitted neural network is then used to compute optimal trading actions, namely, $\boldsymbol{a}(\boldsymbol{X}_t) = \arg\max_a Q(\boldsymbol{X}_t, \boldsymbol{a})$.

## 3.2 Piece-Wise Linear and Exponential Utility Functions

We first test trading signals constructed using a piece-wise linear utility function,

$$U(\boldsymbol{R}) = \max(\boldsymbol{R},\, 0) + \gamma \min(\boldsymbol{R},\, 0)\,, \tag{3.5}$$

and then test using an exponential utility function,

$$U(\boldsymbol{R}) = -\frac{1}{\gamma}\exp(-\gamma \boldsymbol{R})\,, \tag{3.6}$$

where we take the risk aversion coefficient $\gamma = 1.3$ for the piece-wise linear utility function and $\gamma = 3$ for the exponential utility function. We then fit the dense DFN with respect to the piece-wise linear utility function (3.5) with the same quadratic loss given in equation (3.4), and fit the dense DFN with respect to the exponential utility function (3.6) by minimizing the quadratic loss of the certainty equivalent,

$$\min_{\boldsymbol{W},\,\boldsymbol{b}} \sum_{a\in\mathcal{A}} \frac{1}{N}\sum_{i=1}^{N}\left(Q\left(\boldsymbol{X}_0^{(i)},\, \boldsymbol{a}\right) - U^{-1}\left(\frac{1}{M}\sum_{i'=1}^{M} U\left(\boldsymbol{R}_1^{(i,\,i')}(\boldsymbol{a})\right)\right)\right)^2 . \tag{3.7}$$

Figure 3.2 illustrates trading-signal heat plots for the piece-wise linear utility function (3.5) and the exponential utility function (3.6). The most obvious difference is that the piece-wise linear utility has states where the trading signals suggest to take position (0, 0). Table 3.2 and Table 3.3 display the portfolio metrics for the $k$-fold cross validation of out-of-sample tests, and Figure 3.3 shows the time series of portfolio values, as given by the piece-wise linear utility function (3.5) and the exponential utility function (3.6). By observing these tables and figures, strong portfolio performance can be concluded based on the values of profits and Sharpe ratios, but it is important also to highlight the large drawdowns and the difficulty they would pose in practice. In order to perform comprehensive comparisons, we also display some results that are calculated from utilizing benchmarks. Tables B.1 through Table B.5 in Appendix B show similar portfolio metrics for the SPDR S&P 500 Trust ETF, whose ticker symbol is SPY, and the four individual trading actions that are defined in equation (3.1). Notice that none of these benchmarks posts a positive return over every fold, whereas the trading signals given by the neural network do. Moreover, notice in Appendix B that only for fold #1 of constant trading actions $(-1,\, 1)$ and $(-1,\, 2)$ have comparable performance to the results of the neural network; in all other folds there is not any constant trading action choice that is comparable to the trading strategies that are provided by the neural network.

## 3.3 Transaction Costs

Finally, it remains to test if the trading signals from Section 3 can perform with transaction costs. Execution of this strategy is done utilizing market orders, which means that market makers provide liquidity, and therefore, we cross their bid-ask spread each time when we complete a trade. The price data that we use in our backtests are bid-ask midpoints. Thus, to simulate real-life trading of market orders, we should pay (at least) 1/2 the bid-ask spread each time when we open or close a VIX futures position.

VIX futures have a tick size of five cents[6], which means that our backtests should always assume a bid-ask spread of at least five cents in U.S. Dollars. In the simplest backtest, we hold the bid-ask spread constant at five cents, which means we pay \$0.025 each time when we open or close a VIX futures position. However, bid-ask spreads may widen, particularly when the VIX futures curve is in backwardation. With this widening in mind, a transaction-cost function for the $i^{th}$ VIX future is

$$TC_t^i = \frac{1}{2}\max\left(\varepsilon F_t^i,\, 0.05\right) \ , \tag{3.8}$$

where $\varepsilon$ is a fixed basis points (bps) parameter, in other words, $\varepsilon$ is equal to 20bps, 30bps, or 40bps. Using this notation for transaction costs, the returns on the value of the trading-signal portfolio are computed similarly to equation (2.10) except for an additional term for transaction costs,

$$\frac{\Delta P_t}{P_t} = \boldsymbol{R}_{t+1}\left(\boldsymbol{a}\left(\boldsymbol{X}_t\right)\right) + r\Delta t - \frac{1}{P_t}\sum_i TC_t^i\left|n_t^i - n_{t-1}^i\right| \ , \tag{3.9}$$

[6] Each VIX future that is traded on the Chicago Board Options Exchange (CBOE) has a multiplier of 1000, which means that the tick size is effectively 50 U.S. Dollars.

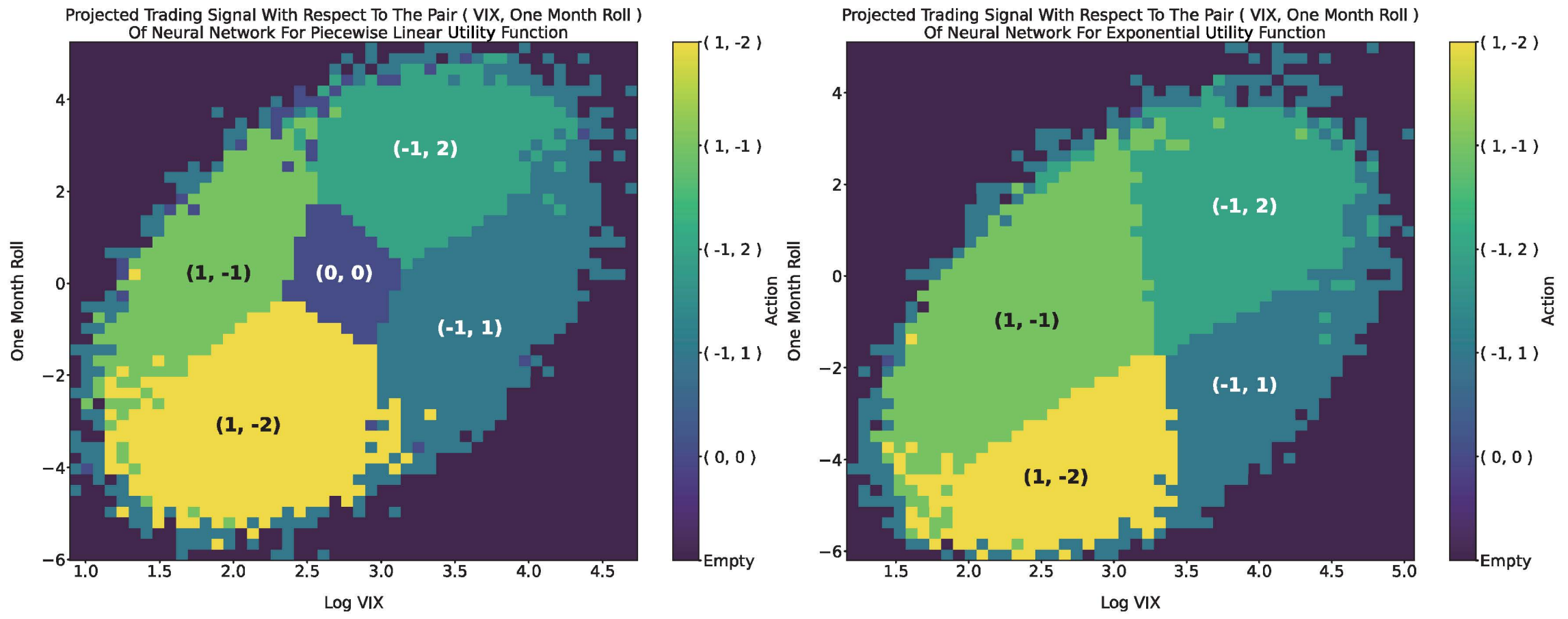

Figure 3.2: Heat plots showing the projection of the trading signals onto the two-dimensional space spanned by logarithm of VIX and the one-month roll, with the projected values being the most-common trading actions at these points. The left plot is the projection of the trading signal constructed with piece-wise linear utility function (3.5), the right plot is the projection of the trading signal constructed with exponential utility function (3.6). Both trading signals are constructed utilizing the dense DFN approach of Section 3.1. The value "Empty" represents the values of logarithm of VIX and one-month roll that do not occur.

| Statistics / Fold | $\mathbb{E}\left[\boldsymbol{R}_t\left(\boldsymbol{a}\left(\boldsymbol{X}_t\right)\right)\right]$ | $\text{std}[\boldsymbol{R}_t\left(\boldsymbol{a}\left(\boldsymbol{X}_t\right)\right)]$ | Profit (%) | Sharpe Ratio | Maximum Drawdown |
|---|---|---|---|---|---|
| 0 | 2.361 | 0.443 | 304.264 | 5.297 | -0.196 |
| 1 | 1.195 | 0.368 | 145.924 | 3.215 | -0.138 |
| 2 | 4.951 | 0.447 | 724.848 | 11.053 | -0.117 |
| 3 | 2.835 | 0.410 | 384.387 | 6.878 | -0.214 |
| 4 | 0.854 | 0.242 | 108.168 | 3.474 | -0.128 |
| 5 | 1.129 | 0.361 | 137.044 | 3.093 | -0.123 |
| 6 | 1.027 | 0.375 | 121.582 | 2.709 | -0.156 |
| 7 | 1.415 | 0.754 | 130.578 | 1.862 | -0.293 |
| 8 | 0.329 | 0.302 | 34.784 | 1.056 | -0.180 |
| 9 | 3.284 | 0.491 | 429.191 | 6.661 | -0.240 |

Table 3.2: Portfolio metrics for out-of-sample tests in $k$-fold cross validation on trading signal constructed with piece-wise linear utility function (3.5). These metrics are computed from the portfolio returns given by equation (2.10) with no transaction costs.

| Fold \ Statistics | $\mathbb{E}\left[\boldsymbol{R}_t\left(\boldsymbol{a}\left(\boldsymbol{X}_t\right)\right)\right]$ | $\mathrm{std}[\boldsymbol{R}_t\left(\boldsymbol{a}\left(\boldsymbol{X}_t\right)\right)]$ | Profit (%) | Sharpe Ratio | Maximum Drawdown |
|---|---|---|---|---|---|
| 0 | 1.728 | 0.456 | 209.177 | 3.763 | -0.239 |
| 1 | 1.610 | 0.412 | 198.810 | 3.880 | -0.165 |
| 2 | 4.534 | 0.462 | 645.731 | 9.775 | -0.175 |
| 3 | 3.138 | 0.463 | 418.006 | 6.751 | -0.202 |
| 4 | 0.742 | 0.284 | 90.055 | 2.577 | -0.190 |
| 5 | 0.683 | 0.388 | 74.484 | 1.731 | -0.236 |
| 6 | 0.886 | 0.400 | 100.162 | 2.187 | -0.189 |
| 7 | 1.239 | 0.785 | 103.799 | 1.564 | -0.281 |
| 8 | 0.639 | 0.356 | 71.310 | 1.765 | -0.171 |
| 9 | 2.518 | 0.564 | 294.004 | 4.443 | -0.248 |

Table 3.3: Portfolio metrics for out-of-sample tests in $k$-fold cross validation with trading signal constructed with exponential utility function (3.6). These metrics are computed from the portfolio returns given by equation (2.10) with no transaction costs.

where $n_t^i$ denotes the number of contracts in the $i^{th}$ VIX future; computation of $n_t^i$ is explained in Appendix A.

Figure 3.4 illustrates the time series of portfolio values for trading signals that are constructed utilizing piece-wise linear utility function (3.5) and utilizing the portfolio values with transaction costs that are given by equation (3.9). Figure 3.5 illustrates the time series of portfolio value for trading signals that are constructed utilizing exponential utility function (3.6) and utilizing the portfolio values with transaction costs that are given by equation (3.9). Table 3.5 and Table 3.4 display the metrics for portfolios that are computed with varying levels of transaction costs utilizing piece-wise linear utility function (3.5) and exponential utility function (3.6). In general, these portfolios can still perform well when transaction costs are included, but we do observe a decline as we increase the basis points parameter in the transaction cost function (3.8). In other words, as we increase the values of $\varepsilon$ in equation (3.8), the profits and Sharpe ratios decrease. Finally, as is the case of no transaction cost, maximum drawdowns remain high when transaction costs are included.

# 4 Conclusion

In this article, we have proposed and analyzed a method for constructing VIX futures trading signals. The basis for the method is in the identification of certain trading opportunities by observing the shape of the VIX futures curve. The trading signal uses a deep feed-forward neural network with dense connective structure to determine the best trading action for day-ahead expectation of returns. We backtested this method and found it to perform well in out-of-sample tests, showing considerable profits and reasonable Sharpe ratios, but also showing levels of maximum drawdown that would be difficult to manage in practice. When we included transaction costs, we observed that the portfolio performance reduced to more pedestrian levels.

# Appendix A Mapping Trading Signal to Futures Positions

The dense DFN model that is proposed in Section 3.1 provides optimal trading actions for the yields of rolling VIX futures strategy $I^i$ described in equation (2.8). For example, the trading signal occurring at the volatility spike on January $26^{th}$ of 2021, the piece-wise linear utility function (3.5) and the deep feed-forward neural network (3.3) produces an optimal action $(-1, 1)$. This represents a position with weight $-1$ in $I^1$ and weight of 1 in $I^5$. This appendix provides a translation of the outputs for the neural network algorithm into the exact quantities that a real-life trader would utilize when setting up a position.

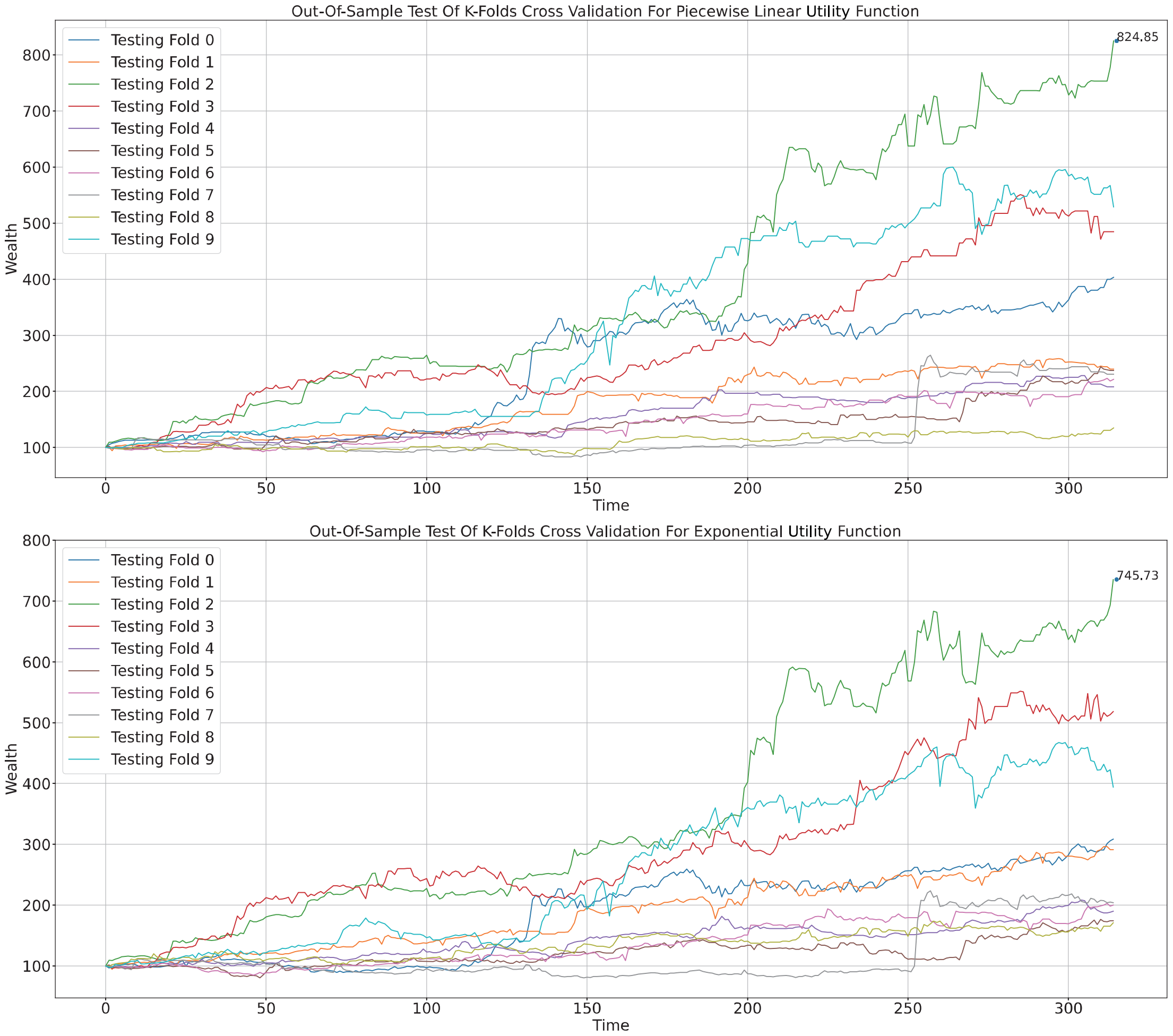

Figure 3.3: Time series of portfolio value for out-of-sample tests of $k$-fold cross validation on trading signals constructed with piece-wise linear utility function (3.5) (top) and exponential utility function (3.6) (bottom). The state-action value function $Q(\boldsymbol{X}_t, \boldsymbol{a})$ is obtained by training the dense DFN given by equation (3.3). The returns $\boldsymbol{R}_t(\boldsymbol{a}(\boldsymbol{X}_t))$ are computed with the trading actions $\boldsymbol{a}(\boldsymbol{X}_t) = \arg\max_{a \in \mathcal{A}} Q(\boldsymbol{X}_t, \boldsymbol{a})$, and the portfolio values are computed from equation (2.10) with no transaction costs.

In expression (3.1), we define the five actions that are considered in our analyses, which are $\boldsymbol{a}_j$ for $j = 0, 1, 2, 3, 4$, with $\boldsymbol{a}_0 = (0, 0)$, $\boldsymbol{a}_1 = (-1, 1)$, $\boldsymbol{a}_2 = (-1, 2)$, $\boldsymbol{a}_3 = (1, -1)$, and $\boldsymbol{a}_4 = (1, -2)$. Each action $\boldsymbol{a}_i$ is a two-dimensional vector,

$$\boldsymbol{a}_j = \left(a_j^1, a_j^5\right)$$

where $a_j^1$ is the portfolio weight for $I^1$ and $a_j^5$ is the weight for $I^5$. This action can be converted into the actual number of contracts in VIX future $F^i$ that is defined by equation (2.1). Letting $n^i$ denote the number of VIX future contract in $F^i$ for $i = 1, 2, 5$, and 6, the followings are the conversions from a given $a^i_j$ to the $n^i$ for trading signals taking positions in $I^1$ and $I^5$,

$$n^1 = \frac{\omega a^1_j P}{V^1}, \qquad n^2 = \frac{(1-\omega)\, a^1_j P}{V^1}, \qquad n^5 = \frac{\omega a^5_j P}{V^5}, \qquad n^6 = \frac{(1-\omega)\, a^5_j P}{V^5}, \tag{A.1}$$

where $P$ denotes the wealth of trading portfolio, given by either equation (2.10) or equation (3.9), and where we have taken the rolling weight $\omega$ to be the same for all $i$ as described in the beginning of Section 3. For example, if the optimal trading action is $(-1, 1)$, then we have $a^1_j = -1$ and $a^5_j = 1$, and we apply accordingly the above equation for $n^i$.

Table A.1 shows the positions in VIX futures $F^i_t$ for a real-time run starting from December $28^{th}$ of 2020 to February $19^{th}$ of 2021 of trading signal constructed with the piece-wise linear utility function (3.5). The portfolio values that are shown in the table includes a transaction cost of 1/2 the bid-ask spread for each trade, in other words, the portfolio values are calculated utilizing equation (3.9) with $\varepsilon = 0$, and each position is rounded to the nearest whole number of contracts. Of note is the drop in $P$ from January $26^{th}$ of 2021 to January $27^{th}$ of 2021, which was the day of the GameStop trading freeze. The trade signal incurred a loss from the VIX spike caused by GameStop, but then recovered the losses in the following days.

# Appendix B Metrics for the SPY and VIX ETFs

Results that are reported in Section 3.1 should be compared with some standard benchmarks and non-neural-network trading actions. Table B.1 presents, for SPDR S&P 500 Trust ETF, the same metrics that are utilized to evaluate the VIX futures trading-signal portfolios for the same ten time periods of the $k$-fold cross validation. Table B.2 to Table B.5 respectively display the same metrics that are utilized to evaluate the VIX futures trading-signal portfolios for the same ten time periods of the $k$-fold cross validation for the trading portfolio utilizing one of the trading actions that are defined in equation (3.1).

By comparing with the metrics for DFN-based trading signals that are displayed in Table 3.2 and Table 3.3 in Section 3.1, we can observe that the results that are produced by the neural network algorithm that we propose have reasonably good returns, profits, and Sharpe ratios, but also have high volatility and high drawdowns.

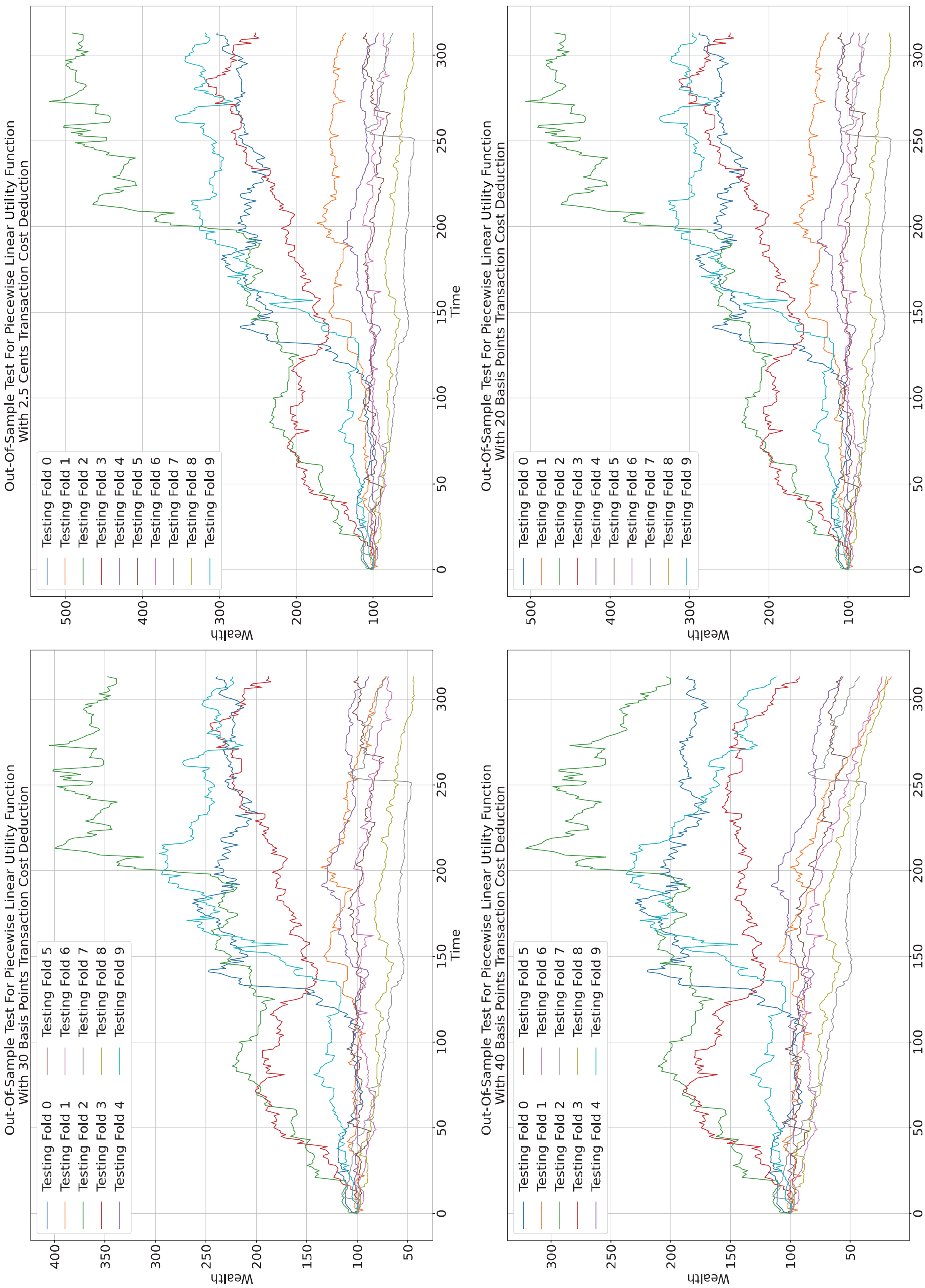

Figure 3.4: Time series of portfolio values that are computed with transaction costs utilizing $\varepsilon = 0$, 20, 30, and 40bps as described in equation (3.8), for out-of-sample tests of $k$-fold cross-validation on trading the signal constructed with piece-wise linear utility function (3.5). The state-action value function $Q(\boldsymbol{X}_t, \boldsymbol{a})$ is obtained by training the dense DFN given by equation (3.3). The returns $\boldsymbol{R}_t(\boldsymbol{a}(\boldsymbol{X}_t))$ are computed with the trading actions $\boldsymbol{a}(\boldsymbol{X}_t) = \arg\max_{a \in \mathcal{A}} Q(\boldsymbol{X}_t, \boldsymbol{a})$, and the portfolio value with transaction cost deduction for the optimal action is $P_t$ computed with equation (3.9).

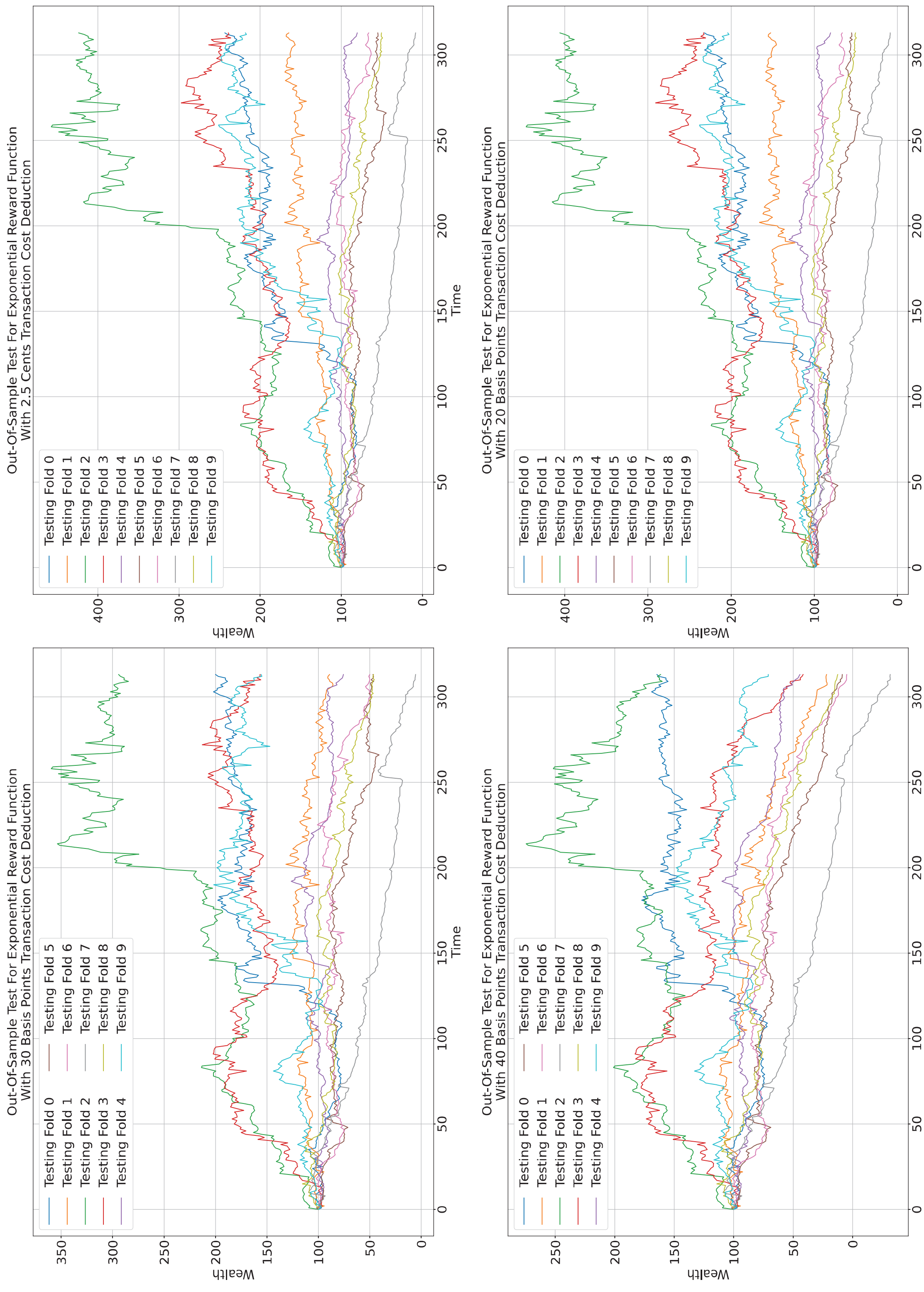

Figure 3.5: Time series of portfolio values that are computed with transaction costs utilizing $\varepsilon = 0$, 20, 30, and 40bps as described in equation (3.8), for out-of-sample tests of $k$-fold cross-validation on trading the signal constructed with exponential utility function (3.6). The state-action value function $Q(\boldsymbol{X}_t, \boldsymbol{a})$ is obtained by training the dense DFN given by equation (3.3). The returns $\boldsymbol{R}_t(\boldsymbol{a}(\boldsymbol{X}_t))$ are computed with the trading actions $\boldsymbol{a}(\boldsymbol{X}_t) = \arg\max_{a \in \mathcal{A}} Q(\boldsymbol{X}_t, \boldsymbol{a})$, and the portfolio value with transaction cost deduction for the optimal action is $P_t$ computed with equation (3.9).

| Statistics \ Fold | | 0 | 1 | 2 | 3 | 4 | 5 | 6 | 7 | 8 | 9 |
|---|---|---|---|---|---|---|---|---|---|---|---|
| $\varepsilon$ =0 | Profit (%) | 202.813 | 35.406 | 391.423 | 152.749 | -6.926 | 8.136 | -14.775 | -26.051 | -52.547 | 218.103 |
| | Sharpe Ratio | 3.756 | 0.956 | 6.643 | 3.099 | -0.159 | 0.345 | -0.175 | -0.049 | -1.403 | 3.768 |
| | Maximum Drawdown | -0.213 | -0.216 | -0.163 | -0.258 | -0.329 | -0.311 | -0.306 | -0.601 | -0.542 | -0.240 |
| | P&L per Transaction | 4.345 | 0.460 | 2.737 | 1.051 | -0.078 | 0.104 | -0.159 | -0.312 | -0.741 | 1.782 |
| $\varepsilon$ = 20bps | Profit (%) | 187.321 | 24.631 | 370.000 | 148.617 | -6.926 | 8.126 | -14.780 | -26.122 | -52.550 | 195.975 |
| | Sharpe Ratio | 3.509 | 0.719 | 6.329 | 3.025 | -0.159 | 0.344 | -0.175 | -0.050 | -1.403 | 3.439 |
| | Maximum Drawdown | -0.225 | -0.262 | -0.163 | -0.262 | -0.329 | -0.311 | -0.306 | -0.601 | -0.542 | -0.240 |
| | P&L per Transaction | 3.176 | 0.287 | 2.306 | 1.012 | -0.078 | 0.104 | -0.159 | -0.313 | -0.741 | 1.403 |
| $\varepsilon$ = 30bps | Profit (%) | 138.855 | -28.295 | 247.105 | 87.970 | -11.656 | -1.441 | -31.083 | -28.788 | -55.441 | 123.111 |
| | Sharpe Ratio | 2.709 | -0.488 | 4.460 | 1.919 | -0.324 | 0.119 | -0.561 | -0.087 | -1.492 | 2.315 |
| | Maximum Drawdown | -0.244 | -0.473 | -0.187 | -0.307 | -0.346 | -0.333 | -0.367 | -0.601 | -0.569 | -0.279 |
| | P&L per Transaction | 1.611 | -0.226 | 1.073 | 0.501 | -0.217 | -0.017 | -0.302 | -0.334 | -0.755 | -0.636 |
| $\varepsilon$ = 40bps | Profit (%) | 85.275 | -84.254 | 103.636 | -7.369 | -43.848 | -42.591 | -76.575 | -57.851 | -80.190 | 11.942 |
| | Sharpe Ratio | 1.787 | -1.677 | 2.096 | 0.046 | -1.458 | -0.898 | -1.658 | -0.509 | -2.170 | 0.456 |
| | Maximum Drawdown | -0.269 | -0.857 | -0.377 | -0.510 | -0.515 | -0.507 | -0.765 | -0.681 | -0.802 | -0.528 |
| | P&L per Transaction | 0.742 | -0.505 | 0.338 | -0.032 | -0.380 | -0.388 | -0.573 | -0.555 | -0.846 | 0.046 |

Table 3.4: Portfolio metrics with transaction cost deduction for out-of-sample tests in $k$-fold cross-validation with trading signals constructed with piece-wise linear utility function (3.5), and the state-action function $Q(\boldsymbol{X}_t, \boldsymbol{a})$ obtained by training the dense DFN given by equation (3.3). The transaction costs per contract are given by equation (3.8) and the portfolio returns are given by equation (3.9).

| Statistics | Fold | 0 | 1 | 2 | 3 | 4 | 5 | 6 | 7 | 8 | 9 |
|---|---|---|---|---|---|---|---|---|---|---|---|
| $\varepsilon = 0$ | Profit (%) | 143.296 | 64.787 | 23.421 | 139.082 | -19.477 | -44.923 | -33.712 | -91.306 | -49.599 | 118.240 |
| | Sharpe Ratio | 2.738 | 1.477 | 5.505 | 2.649 | -0.470 | -0.871 | -0.565 | -0.967 | -1.063 | 2.098 |
| | Maximum Drawdown | -0.274 | -0.184 | -0.224 | -0.263 | -0.381 | -0.563 | -0.443 | -0.920 | -0.574 | -0.328 |
| | P&L per Transaction | 4.595 | 0.819 | 2.413 | 0.843 | -0.232 | -0.545 | -0.348 | -0.854 | -0.523 | 1.169 |
| $\varepsilon = 20$bps | Profit (%) | 131.973 | 52.441 | 305.771 | 133.121 | -19.477 | -44.931 | -33.716 | -91.371 | -49.600 | 103.910 |
| | Sharpe Ratio | 2.553 | 1.239 | 5.245 | 2.551 | -0.470 | -0.871 | -0.565 | -0.968 | -1.063 | 1.892 |
| | Maximum Drawdown | -0.274 | -0.185 | -0.224 | -0.267 | -0.381 | -0.563 | -0.443 | -0.921 | -0.574 | -0.328 |
| | P&L per Transaction | 3.327 | 0.599 | 2.060 | 0.795 | -0.232 | -0.545 | -0.348 | -0.854 | -0.523 | 0.921 |
| $\varepsilon = 30$bps | Profit (%) | 99.883 | -10.819 | 194.296 | 56.765 | -24.365 | -53.164 | -50.059 | -94.346 | -53.466 | 55.666 |
| | Sharpe Ratio | 2.015 | -0.019 | 3.545 | 1.259 | -0.618 | -1.075 | -0.943 | -0.989 | -1.161 | 1.173 |
| | Maximum Drawdown | -0.297 | -0.348 | -0.254 | -0.311 | -0.402 | -0.596 | -0.515 | -0.948 | -0.605 | -0.333 |
| | P&L per Transaction | 1.724 | -0.085 | 0.920 | 0.280 | -0.280 | -0.600 | -0.466 | -0.859 | -0.544 | 0.374 |
| $\varepsilon = 40$bps | Profit (%) | 64.643 | -78.125 | 61.369 | -58.328 | -55.578 | -91.129 | -95.012 | -131.543 | -87.536 | -29.006 |
| | Sharpe Ratio | 1.405 | -1.354 | 1.336 | -0.902 | -1.566 | -1.809 | -1.624 | 0.794 | -1.887 | -0.192 |
| | Maximum Drawdown | -0.330 | -0.808 | -0.416 | -0.772 | -0.596 | -0.912 | -0.950 | -1.289 | -0.890 | -0.534 |
| | P&L per Transaction | 0.837 | -0.460 | 0.218 | -0.219 | -0.503 | -0.792 | -0.683 | -0.986 | -0.690 | -0.147 |

Table 3.5: Portfolio metrics with transaction cost deduction for out-of-sample tests in $k$-fold cross validation with trading signal constructed with exponential utility function (3.6), and the state-action function $Q(\boldsymbol{X}_t, \boldsymbol{a})$ obtained by training the dense DFN given by equation (3.3). The transaction costs per contract are given by equation (3.8) and the portfolio returns are given by equation (3.9).

| Date | $P$ | $\omega$ | $a^1$ | $a^5$ | $n^1$ | $n^2$ | $n^5$ | $n^6$ | $\sum_i n^i$ |
|---|---|---|---|---|---|---|---|---|---|
| 2020-12-28 | 100.00 | 0.65714 | -1 | 2 | -3 | -1 | 5 | 3 | 4 |
| 2020-12-29 | 101.25 | 0.62857 | -1 | 1 | -3 | -1 | 2 | 1 | -1 |
| 2020-12-30 | 103.20 | 0.60000 | 0 | 0 | 0 | 0 | 0 | 0 | 0 |
| 2020-12-31 | 103.03 | 0.57143 | 0 | 0 | 0 | 0 | 0 | 0 | 0 |
| 2021-01-04 | 103.03 | 0.45714 | -1 | 2 | -2 | -2 | 4 | 4 | 4 |
| 2021-01-05 | 102.23 | 0.42857 | -1 | 2 | -2 | -2 | 3 | 4 | 3 |
| 2021-01-06 | 101.15 | 0.40000 | -1 | 2 | -2 | -2 | 3 | 5 | 4 |
| 2021-01-07 | 104.18 | 0.37143 | 0 | 0 | 0 | 0 | 0 | 0 | 0 |
| 2021-01-08 | 103.88 | 0.34286 | -1 | 1 | -1 | -3 | 1 | 3 | 0 |
| 2021-01-11 | 101.23 | 0.25714 | -1 | 1 | -1 | -3 | 1 | 3 | 0 |
| 2021-01-12 | 103.22 | 0.22857 | -1 | 1 | -1 | -3 | 1 | 3 | 0 |
| 2021-01-13 | 105.05 | 0.20000 | -1 | 1 | -1 | -3 | 1 | 3 | 0 |
| 2021-01-14 | 105.45 | 0.17143 | -1 | 1 | -1 | -4 | 1 | 3 | -1 |
| 2021-01-15 | 103.63 | 0.14286 | -1 | 1 | -1 | -3 | 1 | 3 | 0 |
| 2021-01-19 | 105.99 | 0.02857 | -1 | 2 | 0 | -4 | 0 | 8 | 4 |
| 2021-01-20 | 103.19 | 0.00000 | 0 | 0 | 0 | 0 | 0 | 0 | 0 |
| 2021-01-21 | 102.89 | 0.96429 | -1 | 2 | -4 | 0 | 7 | 0 | 3 |
| 2021-01-22 | 103.37 | 0.92857 | 0 | 0 | 0 | 0 | 0 | 0 | 0 |
| 2021-01-25 | 103.10 | 0.82143 | -1 | 1 | -3 | -1 | 3 | 1 | 0 |
| 2021-01-26 | 105.44 | 0.78571 | -1 | 1 | -3 | -1 | 3 | 1 | 0 |
| 2021-01-27 | 92.60 | 0.75000 | -1 | 2 | -2 | -1 | 5 | 2 | 4 |
| 2021-01-28 | 90.38 | 0.71429 | -1 | 2 | -2 | -1 | 4 | 2 | 3 |
| 2021-01-29 | 89.03 | 0.67857 | -1 | 2 | -2 | -1 | 4 | 2 | 3 |
| 2021-02-01 | 92.25 | 0.57143 | -1 | 2 | -2 | -1 | 4 | 3 | 4 |
| 2021-02-02 | 93.27 | 0.53571 | -1 | 2 | -2 | -2 | 3 | 3 | 2 |
| 2021-02-03 | 96.28 | 0.50000 | 0 | 0 | 0 | 0 | 0 | 0 | 0 |
| 2021-02-04 | 96.03 | 0.46429 | -1 | 1 | -2 | -2 | 2 | 2 | 0 |
| 2021-02-05 | 96.69 | 0.42857 | -1 | 1 | -2 | -2 | 1 | 2 | -1 |
| 2021-02-08 | 98.28 | 0.32143 | -1 | 1 | -1 | -3 | 1 | 2 | -1 |
| 2021-02-09 | 98.66 | 0.28571 | -1 | 1 | -1 | -3 | 1 | 2 | -1 |
| 2021-02-10 | 98.85 | 0.25000 | -1 | 1 | -1 | -3 | 1 | 3 | 0 |
| 2021-02-11 | 100.93 | 0.21429 | -1 | 1 | -1 | -3 | 1 | 3 | 0 |
| 2021-02-12 | 104.66 | 0.17857 | -1 | 1 | -1 | -3 | 1 | 3 | 0 |
| 2021-02-16 | 105.57 | 0.03571 | 1 | -1 | 0 | 4 | 0 | -3 | 1 |
| 2021-02-17 | 115.42 | 0.00000 | -1 | 1 | 0 | -4 | 0 | 4 | 0 |
| 2021-02-18 | 115.25 | 0.96429 | -1 | 1 | -4 | 0 | 4 | 0 | 0 |
| 2021-02-19 | 118.17 | 0.92857 | 0 | 0 | 0 | 0 | 0 | 0 | 0 |

Table A.1: Real-time backtest results from December $28^{th}$ of 2020 to February $19^{th}$ of 2021 for the piece-wise linear utility function (3.5). The dense feed-forward neural network (3.3) is re-trained weekly, the numbers of contracts $n^i$ in $F^i_t$ are given by equation (A.1), the portfolio value $P$ is given in equation (3.9) with $\varepsilon = 0$, the CMF roll weight $\omega$ appears in equation (2.2), and $i$ is explained in Section 3. The net position in VIX futures is $\sum_i n^i$. The position is long if the net position is positive, short if the net position is negative, and neutral if the net position is zero. Of note is the loss observed from Jan. 26th to 27th during the GameStop trading freeze, and then recovery of losses in the following days.

| Fold \ Statistics | $\mathbb{E}\left[\boldsymbol{R}_t\left(\boldsymbol{a}\left(\boldsymbol{X}_t\right)\right)\right]$ | $\text{std}[\boldsymbol{R}_t\left(\boldsymbol{a}\left(\boldsymbol{X}_t\right)\right)]$ | Profit (%) | Sharpe Ratio | Maximum Drawdown |
|---|---|---|---|---|---|
| 0 | -0.172 | 0.415 | -29.111 | -0.439 | -0.514 |
| 1 | 0.216 | 0.180 | 25.147 | 1.144 | -0.157 |
| 2 | 0.138 | 0.211 | 14.347 | 0.607 | -0.186 |
| 3 | 0.192 | 0.125 | 23.383 | 1.452 | -0.096 |
| 4 | 0.222 | 0.105 | 27.567 | 2.014 | -0.056 |
| 5 | 0.100 | 0.144 | 11.191 | 0.624 | -0.119 |
| 6 | 0.097 | 0.130 | 11.171 | 0.676 | -0.128 |
| 7 | 0.155 | 0.114 | 18.722 | 1.271 | -0.101 |
| 8 | 0.095 | 0.145 | 10.514 | 0.583 | -0.193 |
| 9 | 0.218 | 0.304 | 20.637 | 0.685 | -0.337 |

Table B.1: Metrics for SPDR S&P 500 Trust ETF (ticker symbol: SPY) for the same ten folds listed in Table 3.1, which are utilized in the $k$-fold cross validation of Section 3.

| Fold \ Statistics | $\mathbb{E}\left[\boldsymbol{R}_t\left(\boldsymbol{a}\left(\boldsymbol{X}_t\right)\right)\right]$ | $\text{std}[\boldsymbol{R}_t\left(\boldsymbol{a}\left(\boldsymbol{X}_t\right)\right)]$ | Profit (%) | Sharpe Ratio | Maximum Drawdown |
|---|---|---|---|---|---|
| 0 | -0.199 | 0.461 | -26.908 | -0.453 | -0.597 |
| 1 | 1.493 | 0.445 | 141.464 | 3.334 | -0.298 |
| 2 | 0.059 | 0.520 | -7.700 | 0.093 | -0.558 |
| 3 | 0.285 | 0.563 | 9.509 | 0.488 | -0.334 |
| 4 | 0.246 | 0.347 | 17.518 | 0.679 | -0.249 |
| 5 | -0.011 | 0.458 | -10.867 | -0.045 | -0.363 |
| 6 | 0.729 | 0.505 | 54.305 | 1.421 | -0.290 |
| 7 | -0.156 | 0.802 | -48.498 | -0.207 | -0.795 |
| 8 | 0.126 | 0.423 | 2.731 | 0.274 | -0.364 |
| 9 | 0.320 | 0.603 | 9.142 | 0.513 | -0.645 |

Table B.2: Portfolio metrics of fixed trading strategy $(-1,\ 1)$ for the same ten folds listed in Table 3.1 without transaction costs, which are utilized in the $k$-fold cross validation of Section 3.

| Fold \ Statistics | $\mathbb{E}\left[\boldsymbol{R}_t\left(\boldsymbol{a}\left(\boldsymbol{X}_t\right)\right)\right]$ | $\text{std}[\boldsymbol{R}_t\left(\boldsymbol{a}\left(\boldsymbol{X}_t\right)\right)]$ | Profit (%) | Sharpe Ratio | Maximum Drawdown |
|---|---|---|---|---|---|
| 0 | 0.196 | 0.427 | 9.200 | 0.436 | -0.287 |
| 1 | 1.467 | 0.374 | 146.905 | 3.902 | -0.120 |
| 2 | -0.113 | 0.354 | -16.253 | -0.349 | -0.362 |
| 3 | -0.307 | 0.415 | -34.318 | -0.763 | -0.574 |
| 4 | -0.104 | 0.249 | -12.872 | -0.459 | -0.331 |
| 5 | 0.043 | 0.264 | 0.721 | 0.125 | -0.225 |
| 6 | 0.379 | 0.330 | 31.456 | 1.120 | -0.166 |
| 7 | -0.361 | 0.574 | -44.803 | -0.647 | -0.637 |
| 8 | 0.281 | 0.267 | 24.143 | 1.015 | -0.151 |
| 9 | 1.096 | 0.537 | 88.277 | 2.020 | -0.313 |

Table B.3: Portfolio metrics of fixed trading strategy $(-1,\ 2)$ for the same ten folds listed in Table 3.1 without transaction costs, which are utilized in the $k$-fold cross validation of Section 3.

| Fold \ Statistics | $\mathbb{E}\left[\boldsymbol{R}_t\left(\boldsymbol{a}\left(\boldsymbol{X}_t\right)\right)\right]$ | $\mathrm{std}[\boldsymbol{R}_t\left(\boldsymbol{a}\left(\boldsymbol{X}_t\right)\right)]$ | Profit (%) | Sharpe Ratio | Maximum Drawdown |
|---|---|---|---|---|---|
| 0 | 0.273 | 0.461 | 14.908 | 0.571 | -0.361 |
| 1 | -0.592 | 0.445 | -56.353 | -1.354 | -0.734 |
| 2 | -0.036 | 0.520 | -15.183 | -0.089 | -0.386 |
| 3 | -0.206 | 0.563 | -30.512 | -0.384 | -0.474 |
| 4 | -0.181 | 0.347 | -22.036 | -0.551 | -0.366 |
| 5 | 0.031 | 0.457 | -6.840 | 0.046 | -0.407 |
| 6 | -0.410 | 0.505 | -44.223 | -0.832 | -0.689 |
| 7 | 0.208 | 0.802 | -6.190 | 0.247 | -0.406 |
| 8 | -0.094 | 0.423 | -16.518 | -0.246 | -0.426 |
| 9 | -0.227 | 0.603 | -33.219 | -0.939 | -0.661 |

Table B.4: Portfolio metrics of fixed trading strategy $(1, -1)$ for the same ten folds listed in Table 3.1 without transaction costs, which are utilized in the $k$-fold cross validation of Section 3.

| Fold \ Statistics | $\mathbb{E}\left[\boldsymbol{R}_t\left(\boldsymbol{a}\left(\boldsymbol{X}_t\right)\right)\right]$ | $\mathrm{std}[\boldsymbol{R}_t\left(\boldsymbol{a}\left(\boldsymbol{X}_t\right)\right)]$ | Profit (%) | Sharpe Ratio | Maximum Drawdown |
|---|---|---|---|---|---|
| 0 | -0.147 | 0.427 | -21.336 | -0.368 | -0.437 |
| 1 | -0.588 | 0.374 | -55.217 | -1.601 | -0.721 |
| 2 | 0.150 | 0.354 | 8.118 | 0.397 | -0.217 |
| 3 | 0.471 | 0.415 | 36.258 | 1.111 | -0.246 |
| 4 | 0.139 | 0.249 | 10.495 | 0.517 | -0.180 |
| 5 | -0.022 | 0.264 | -5.455 | -0.121 | -0.206 |
| 6 | -0.260 | 0.330 | -28.469 | -0.821 | -0.473 |
| 7 | 0.596 | 0.574 | 40.017 | 1.022 | -0.250 |
| 8 | -0.204 | 0.267 | -22.303 | -0.801 | -0.318 |
| 9 | -0.514 | 0.537 | -52.529 | -0.976 | -0.691 |

Table B.5: Portfolio metrics of fixed trading strategy $(1, -2)$ for the same ten folds listed in Table 3.1 without transaction costs, which are utilized in the $k$-fold cross validation of Section 3.